\documentclass[aps,twocolumn,amssymb,floatfix,preprintnumbers,amsmath,amssymb, prb]{revtex4-1}
\usepackage{epsfig}
\usepackage{graphicx}%
\usepackage{epstopdf}
\usepackage{dcolumn}
\usepackage{bm}
\usepackage{ifthen}
\usepackage{booktabs}
\usepackage{SIunits}
\usepackage{ulem}
\usepackage{xcolor}

\begin{document}

\title{Detection of defect-induced magnetism in low-dimensional ZnO structures by Magnetophotocurrent}
\author{Israel Lorite}
 \affiliation{Division of Superconductivity and Magnetism, Institut f\"ur Experimentelle Physik II, Fakult\"{a}t f\"{u}r Physik und Geowissenschaften, Universit\"at Leipzig, Linn\'{e}strasse 5, 04103 Leipzig, Germany}
\author{Yogesh Kumar}
 \affiliation{Division of Superconductivity and Magnetism, Institut f\"ur Experimentelle Physik II, Fakult\"{a}t f\"{u}r Physik und Geowissenschaften, Universit\"at Leipzig, Linn\'{e}strasse 5, 04103 Leipzig, Germany}
\author{Pablo Esquinazi}
 \affiliation{Division of Superconductivity and Magnetism, Institut f\"ur Experimentelle Physik II, Fakult\"{a}t f\"{u}r Physik und Geowissenschaften, Universit\"at Leipzig, Linn\'{e}strasse 5, 04103 Leipzig, Germany}
\author{Carlos Zandalazini}
 \affiliation{Laboratorio de F\'{i}sica del S\'{o}lido,  Dpto. de
F\'{i}sica\\ Facultad de Ciencias Exactas y Tecnolog\'{i}a,
Universidad Nacional de Tucum\'{a}n, Argentina}
\author{Silvia Perez de Heluani}
 \affiliation{Laboratorio de F\'{i}sica del S\'{o}lido,  Dpto. de
F\'{i}sica\\ Facultad de Ciencias Exactas y Tecnolog\'{i}a,
Universidad Nacional de Tucum\'{a}n, Argentina}
\date{\today}

\begin{abstract}

The detection of defect-induced  magnetic order in single
low-dimensional oxide structures is in general  difficult  because
of the relatively small yield of magnetically ordered regions. In
this work we have studied the effect of an external magnetic field
on the transient photocurrent measured after light irradiation on
different ZnO samples at room temperature. We found that a
magnetic field produces a change in the relaxation rate of the
transient photocurrent only in magnetically ordered ZnO samples.
This rate can decrease or increase with field depending whether
the magnetic order region is in the bulk or only at the surface of
the ZnO sample. The phenomenon reported here is of importance for
the development of magneto-optical low-dimensional  oxides devices
and
 provide a new guideline for the detection of magnetic order in
 low-dimensional magnetic semiconductors.

\end{abstract}

\pacs{BUSCAR PACS}
\maketitle


The detection of low levels of magnetic moment  is of importance
for the study of new materials and new phenomena in solid state
magnetism.  New technologies such as the simultaneous use of a
superconducting quantum interferometer device (SQUID) and the Moke
effect, are being developed to detect low  levels of magnetic
signals \cite{RiceNAt}. This development is of particular interest
for the study of materials that show magnetic order at
surprisingly high temperatures due to a relatively high
concentration of defects, like vacancies \cite{xin11}. Defect-induced
magnetism (DIM) in semiconducting materials, especially oxides, is of
interest  also because of the expected advantages one has combining
optical and semiconducting properties for spintronics
applications. However, at present there are several difficulties,
partially in the  reproducibility of the DIM due to the unknown
systematic preparation of the necessary defects density, and
also because of  the low levels of magnetic signal due to  the
small amount of magnetically ordered regions in a sample.\cite{abraham05,jag06}

It has recently been shown that the photocurrent is sensitive to
the minority carriers at the interphase between two semiconductors
and it can be changed by an external magnetic field,   opening new
possibilities to design new devices based on photocurrent
processes and interphases properties \cite{Zheng}. On the other
hand, it has been observed that the photocurrent can be affected
by relatively low applied fields of the order of 0.5~T or less, in
diluted magnetic semiconductors at room temperature thanks to the
exchange interaction \cite{Lung}. However, little attention
has been drawn to persistent photocurrent phenomena and their
sensitivity to the variation of the oxygen absorption  after
photoexcitation \cite{zap11,Cai}. Such photocurrent phenomena
are of interest to study  the effect of external magnetic fields
on photo-transport mechanisms  at the surface and/or in bulk of
defect-induced magnetic oxides.

There is consent nowadays that the photocurrent in ZnO is mainly a
surface effect \cite{Park}. The model generally used to understand
photocurrent effects is based on the  desorption and absorption of
oxygen at the ZnO surface. Oxygen from the environment are
chemisorbed at the surface and are negatively charged by capturing
a free carrier, producing at the same time a depletion zone or
surface band bending \cite{MPheno}. During UV-illumination
electron-holes are created. The electrons are promoted to the
conduction band and the holes from valence band and/or intraband
donor defects such as  oxygen vacancies (V$_{\rm O}$) and O
interstitials (O$_{i}$) \cite{MPheno}, can migrate to the surface
to discharge the oxygen at the surface. This produces its
photodesorption and  decreases the band bending,  allowing the
free carrier to diffuse in both directions, i.e. from bulk to
surface and vice versa. After turning off the light excitons are
recombined and the oxygen chemisorbed  at the surface captures a
free carrier,  reducing  the free carrier density and reestablish
finally the initial dark state.

The oxygen reabsorption at the surface provides the main
contribution that affects the transient photocurrent. It depends
on the environmental oxygen concentration and inversely on the
rate of recombination of electrons at traps centres
\cite{Spencer}. In this work we studied in detail the effect of a
magnetic field on the transient photocurrent produced after
turning off the light. The presented results indicate that
the influence of an external magnetic
field on the transient photocurrent is observed only in  samples
with defect-induced magnetic order. The effects are  detectable
thanks to the sensitivity of the phototransport processes. They
help to detect the existence of magnetic order at high
temperatures in low-dimensional ZnO structures, as  single
nano/microwires or microstructured thin films, which would remain
undetectable  by standard techniques.

\section{Method}

For the photocurrent measurements we used four different ZnO
samples. Three samples were 300~nm thin films grown on
 6 $\times$ 6 mm$^{2}$ \textit{a}-plane Al$_{2}$O$_{3}$
substrates by Pulsed Lased Deposition (PLD). Two different films
were grown, one in oxygen  atmosphere (ZO) and another in nitrogen (ZN)
enviroment with partial pressures of $1.73 \times 10^{-2}$~mbar from a base vacuum of
10$^{-7}$~mbar \cite{kha09}. The ZO film was treated afterwards
with hydrogen plasma (ZOH) ($\approx$ 10$^{18}$ H$^{+}$/cm$^{2}$
at an energy of 300~eV at room temperature) \cite{kha11}. All
samples were patterned to a size of $\sim 10 \times 5~\mu$m$^2$ by
e-beam lithography and wet-etching techniques \cite{barpat}.

The fourth sample was a single ZnO microwire with a diameter of
$10~\mu$m and $200~\mu$m length. The wire was doped
with Li (5\%) and treated with hydrogen plasma (ZLH) in a similar
way. Previous studies revealed that the room temperature
hydrogen-plasma treatment in Li-doped ZnO samples
(Li-concentration $\geq 3\%$) produces Zn-vacancies (V$_{\rm Zn}$)
in 10~nm depth from the wire surface. The used
procedure produces  a V$_{\rm Zn}$ concentration similar to the Li
one and magnetic order can be observed at room temperature, as
revealed by XMCD and SQUID techniques \cite{lor15}.

The magneto-photocurrent measurements were performed in a home
made close-cycle cryostat with the possibility to apply a magnetic
field of 0.4~T. The field was always applied parallel to the input
current direction, also parallel to the main film area or the main
axis of the microwire. During the measurements the samples were
kept always at 5x10$^{-3}$~mbar to ensure a constant oxygen
partial pressure at 305~K. To prove the importance of the oxygen
partial pressure in the persistent photocurrent relaxation, we
performed similar experiments after purging the chamber with He to
reduce the partial pressure of oxygen. The reduction of the oxygen
partial pressure in the chamber increased the photocurrent
relaxation time by several hours, in agreement with previous
reports \cite{Hou}. It points out the importance of the oxygen
absorption for the persistent photocurrent process. The light
irradiation was done using a Xenon lamp coupled to a monochromator
from which we selected a wavelength of $\lambda = 370~$nm for all
experiments.
\begin{figure}
     \includegraphics[width=\linewidth]{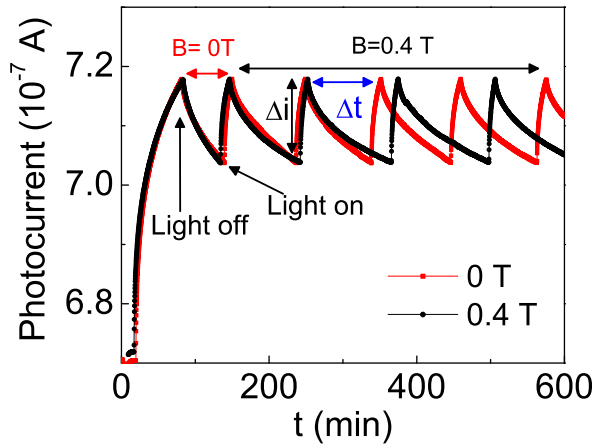}
\caption{\label{Fig1} Experimental  method to measure the effect
of an external magnetic field on the transient photocurrent. The
two curves represent the cycles measured at $B = 0~$T and
$B=0.4$~T of ZN thin film}
\end{figure}

\section{Results}
 The methodology used to obtain the change of the transient photocurrent
 on time and under a magnetic field consists of measuring
consecutively several light-on/off cycles of the
photocurrent. In such an experiment the samples are illuminated
until the photocurrent increases  to a previously selected value
$i_{\rm ON}$ different from saturation. After reaching it, the
light is turned off and the transient photocurrent decreases. When
it reaches a previously selected value $i_{\rm OFF}$, see
Fig.~\ref{Fig1}, the light is turned on till the photocurrent
reaches $i_{\rm ON}$. The time $\Delta t$ for every transient
photocurrent cycle was measured for a previously fixed
$\Delta\textit{i}=i_{\rm ON}-i_{\rm OFF}$. The same procedure is repeated several times to
obtain the time dependence of the transient photocurrent rate
(TPR)  defined as $\frac{\Delta i}{\Delta t}(t)$. The first cycle
performed in every measurement is done at zero magnetic field
($B$) to ensure an equal starting state for every measurement of
the same sample. Note that the selected $\Delta i$ depends on the
sample because the photocurrent varies for every sample.

\begin{figure}
    \includegraphics[width=0.9\linewidth]{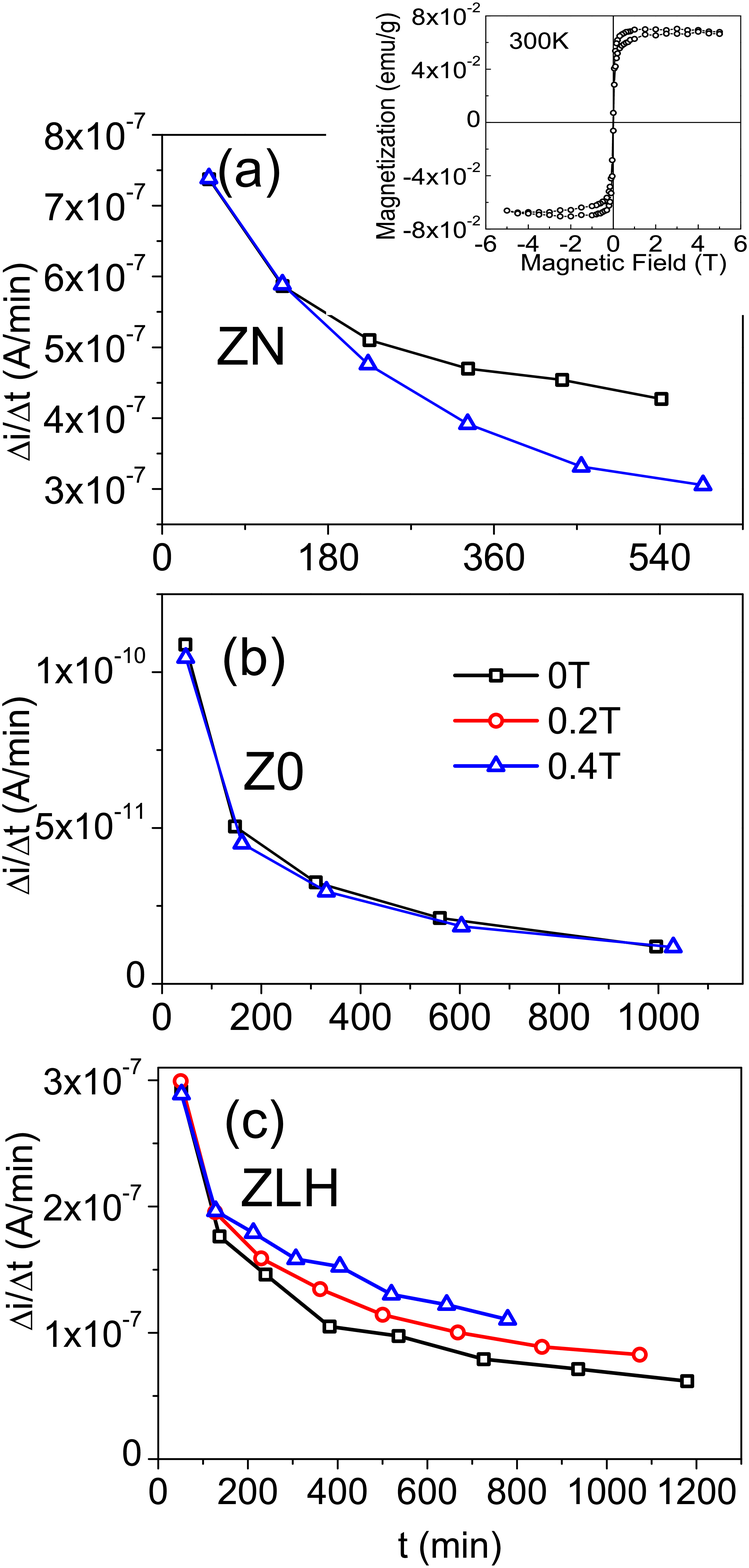}
\caption{\label{Fig2} Variation of the transient photocurrent rate
at different applied magnetic fields, as a function of the total
time for the thin film samples (a)ZN and (b) ZO and the single
microwire (c) ZLH. All measurements were done
at a fixed temperature of 305~K. The inset in (a) shows the field
loop of the magnetization measured for the ZN film at 300~K.}
\end{figure}

To check the equipment performance and the reproducibility of
 previously published variation of the TPR with
 magnetic field \cite{zap11}, we have measured the ZN film, which shows
 magnetic order at room temperature due to the
 existence of V$_{\rm Zn}$ produced during preparation (see inset in
 Fig. \ref{Fig2}(a)) \cite{kha09}. The cycles measured for the ZN sample at zero and
0.4~T field are shown in Fig.~\ref{Fig1}. For both fields we
observe that there is an increase of $\Delta$\textit{t} after
every cycle. This is due to an increase of the number of traps(holes)
created during photoexcitation \cite{Pavel}.  Since not all the
holes are recombined during the light is off an additional
number of photo-generated holes are produced every cycle. The larger the number of
holes the larger the exciton recombination rate, which implies
that the number of carriers moving from bulk to surface, necessary
to initiate the oxygen reabsorption and reestablish the initial
dark conditions, reduces. Thus, the TPR decreases with time, see
Fig.~\ref{Fig2}(a).

When an external magnetic field is applied, we observe
qualitatively a similar variation for the TPR to the one observed
at zero field, see Fig.~\ref{Fig2}(a), however, the variation of $\Delta$\textit{t}
is larger. It means there is an increase of the transient
photocurrent time due to the effect of the external magnetic
field, which should be related to a reduction of the oxygen
absorption rate in comparison to the zero field state. The observed behavior is in
good agreement with that reported previously \cite{zap11}.

After observing the magnetic field effect on the TPR of a magnetic sample,
a ZO thin film was measured, which does not show any sign of magnetic order at room
temperature. The TPR(t) results shown in Fig.~\ref{Fig2}(b)
indicates that there is no dependence with the external magnetic field.

To gain a deeper insight into the field dependence of the TPR
 we compare the obtained
results with those from a sample that shows magnetic order only
within a surface shell, instead of a bulk magnetic sample such as
in ZN. The implantation of $H^{+}$ in the ZLH sample produces a
significant amount of defects within a 10~nm surface
region \cite{kha11,lor15}. Different experimental methods
including XMCD as well as numerical calculations indicate that the
produced concentration of $V_{\rm Zn}$ ($\sim 5\%$) near the Li ions
is the reason for the magnetic order at room temperature
\cite{lor15}. The results for ZLH are shown in Fig.~\ref{Fig2}(c).
In contrast to ZN, the TPR increases with magnetic field.

\begin{figure}
     \includegraphics[width=\linewidth]{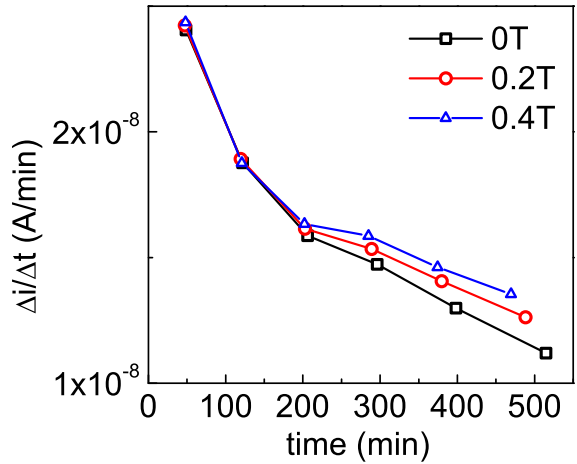}
\caption{\label{Fig4}  Time dependence of the transient
photocurrent rate for ZOH film, at different
applied magnetic fields at 305~K.}
\end{figure}

Because the results indicate that the TPR can depend on the applied magnetic field
only for magnetically ordered samples, we implanted 10$^{18}$ H$^{+}$/cm$^{2}$ in the non-magnetic ZO thin
film shown in Fig.~\ref{Fig2}(b), labeled now ZOH. Such implantation
of H$^{+}$ can produce a near surface
magnetic layer within $\simeq 10$~nm, similarly to the one
observed in ZnO single crystals \cite{kha11}. The results of
the TPR for this sample are shown in Fig.~\ref{Fig4}. The observed
variation with field is similar to the one observed for the ZLH
microwire suggesting that its origin should be related to the
formation of a magnetically surface shell. It is worth to note
that the amount of magnetically ordered mass in this sample is too
small to be measurable with a commercial SQUID, i.e. its
ferromagnetic moment at saturation is smaller than $2 \times
10^{-7}$~emu. Nevertheless, the photocurrent is influenced by the
thin magnetic layer emphasizing the sensitivity of this property
to magnetic order produced by defects in this case.

To understand qualitatively the observed phenomenon we take into
account  the  role of the induced magnetic order in the bulk and
 at the near surface region. In both cases we have the Zeeman splitting
produced by the finite magnetic field, as consequence of the spin-orbit ($L-S$)
coupling with the magnetic defects, $V_{\rm Zn}$, which is large
 for wide band gap magnetic semiconductors
 \cite{Claus}. Due to optical selection rules, which accompany the Zeeman splitting
 and where the spin
orientation of the magnetic defects plays a role \cite{Vece}, the
trapping probability of conduction electrons by the defects
increases with field. This effect reduces the amount of electrons
that can reach the surface and capture oxygen from the
environment to reestablished the initial dark conditions. In this
case the TPR decreases with applied magnetic field, see
Fig.~\ref{Fig2}(a).

\begin{figure}
      \includegraphics[width=\linewidth]{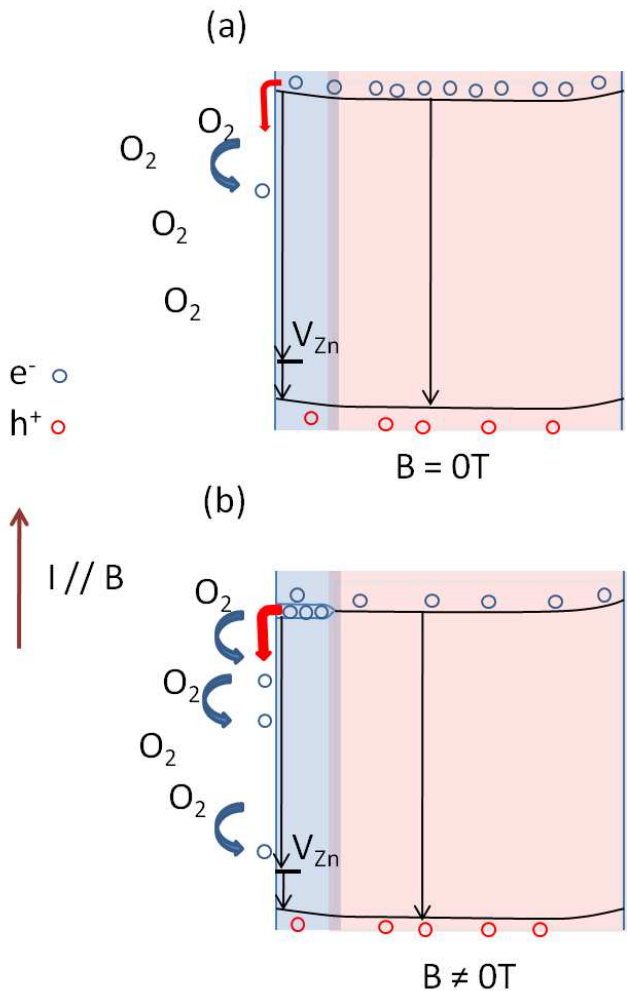}
\caption{\label{Fig3} Sketch of the magnetically ordered
near surface region of a low-energy H$^+$ implanted ZnO sample,
at $B=0~$T (a) and at B$\neq$0~T (b). The magnetic
field is applied parallel to the input current $I$. The blue surface
region of thickness $\sim 10~$nm, denotes the defect-induced magnetically order region and the
pink part is the non-magnetic bulk region. Under a magnetic field
the magnetic region shows a band splitting due to the Zeeman
effect, changing the probability of carrier accumulation at the
surface and influencing the transient photocurrent rate. The Zeeman splitting is not in scale with the
drawn energy gap.}
\end{figure}

In the second case, observed for samples ZLH and ZOH, we must
consider the  heterostructure with only the near surface
region magnetically ordered, see Fig.~\ref{Fig3}. When an external
magnetic field is applied, similarly to the first case, a Zeeman
splitting occurs in the magnetic region. This energy shift
 leads to a
redistribution of carriers with larger probability to be at the
deeper potential. In this case and just after removal of the light
excitation, a larger accumulation of electrons occurs at the
magnetic region due to the carriers coming from the non-magnetic
bulk region. Therefore, more electrons reach the surface
increasing the probability for oxygen trapping at the surface and
as consequence a faster TPR \cite{Yasu} is observed.

The presented results clearly indicate a dependence of the TPR
with relatively low magnetic fields at room temperature. This magnetic field
dependence is qualitatively different for ZnO samples with
magnetic order in the bulk or at the near surface region. The
observed change in the TPR is related to the variation of the rate
of oxygen absorption to reach the initial dark state. This oxygen
absorption rate depends on the number of free electrons reaching the surface, and their
 amount depends on magnetic field and
the magnetically ordered region in the sample. Our results
indicate that low fields can be used to tune the transient
photocurrent in magnetic samples, which can be useful to detect
sensitively room temperature magnetic order in low-dimensional
systems, a magnetic order that may remain undetectable using
routine techniques such as SQUID.

\begin{acknowledgments}
We thank W. Hergert for fruitful discussions.    This  work was
funded by the DFG through the Collaborative Research Center
SFB~762 ``Functionality of Oxide Interfaces'' in Germany, and by
the Secyt-UNCor, CIUNT and Conicet in Argentina.
\end{acknowledgments}

\bibliographystyle{apsrev4-1}

\begin{thebibliography}{20}%
\makeatletter
\providecommand \@ifxundefined [1]{%
 \@ifx{#1\undefined}
}%
\providecommand \@ifnum [1]{%
 \ifnum #1\expandafter \@firstoftwo
 \else \expandafter \@secondoftwo
 \fi
}%
\providecommand \@ifx [1]{%
 \ifx #1\expandafter \@firstoftwo
 \else \expandafter \@secondoftwo
 \fi
}%
\providecommand \natexlab [1]{#1}%
\providecommand \enquote  [1]{``#1''}%
\providecommand \bibnamefont  [1]{#1}%
\providecommand \bibfnamefont [1]{#1}%
\providecommand \citenamefont [1]{#1}%
\providecommand \href@noop [0]{\@secondoftwo}%
\providecommand \href [0]{\begingroup \@sanitize@url \@href}%
\providecommand \@href[1]{\@@startlink{#1}\@@href}%
\providecommand \@@href[1]{\endgroup#1\@@endlink}%
\providecommand \@sanitize@url [0]{\catcode `\\12\catcode
`\$12\catcode
  `\&12\catcode `\#12\catcode `\^12\catcode `\_12\catcode `\%12\relax}%
\providecommand \@@startlink[1]{}%
\providecommand \@@endlink[0]{}%
\providecommand \url  [0]{\begingroup\@sanitize@url \@url }%
\providecommand \@url [1]{\endgroup\@href {#1}{\urlprefix }}%
\providecommand \urlprefix  [0]{URL }%
\providecommand \Eprint [0]{\href }%
\providecommand \doibase [0]{http://dx.doi.org/}%
\providecommand \selectlanguage [0]{\@gobble}%
\providecommand \bibinfo  [0]{\@secondoftwo}%
\providecommand \bibfield  [0]{\@secondoftwo}%
\providecommand \translation [1]{[#1]}%
\providecommand \BibitemOpen [0]{}%
\providecommand \bibitemStop [0]{}%
\providecommand \bibitemNoStop [0]{.\EOS\space}%
\providecommand \EOS [0]{\spacefactor3000\relax}%
\providecommand \BibitemShut  [1]{\csname bibitem#1\endcsname}%
\let\auto@bib@innerbib\@empty
\bibitem [{\citenamefont {Rice}\ \emph {et~al.}(2014)\citenamefont {Rice},
  \citenamefont {Ambwani}, \citenamefont {Bombeck}, \citenamefont {Thompson},
  \citenamefont {Haugstad}, \citenamefont {Leighton},\ and\ \citenamefont
  {Crooker}}]{RiceNAt}%
  \BibitemOpen
  \bibfield  {author} {\bibinfo {author} {\bibfnamefont {W.~D.}\ \bibnamefont
  {Rice}}, \bibinfo {author} {\bibfnamefont {P.}~\bibnamefont {Ambwani}},
  \bibinfo {author} {\bibfnamefont {M.}~\bibnamefont {Bombeck}}, \bibinfo
  {author} {\bibfnamefont {J.~D.}\ \bibnamefont {Thompson}}, \bibinfo {author}
  {\bibfnamefont {G.}~\bibnamefont {Haugstad}}, \bibinfo {author}
  {\bibfnamefont {C.}~\bibnamefont {Leighton}}, \ and\ \bibinfo {author}
  {\bibfnamefont {S.~A.}\ \bibnamefont {Crooker}},\ }\href@noop {} {\bibfield
  {journal} {\bibinfo  {journal} {Nature Materials}\ }\textbf {\bibinfo
  {volume} {13}},\ \bibinfo {pages} {481} (\bibinfo {year} {2014})}\BibitemShut
  {NoStop}%
\bibitem [{\citenamefont {Xing}\ \emph {et~al.}(2011)\citenamefont {Xing},
  \citenamefont {Lu}, \citenamefont {Tian}, \citenamefont {Yi}, \citenamefont
  {Lim}, \citenamefont {Li}, \citenamefont {Li}, \citenamefont {Wang},
  \citenamefont {Yao}, \citenamefont {Ding},\ and\ \citenamefont
  {Feng}}]{xin11}%
  \BibitemOpen
  \bibfield  {author} {\bibinfo {author} {\bibfnamefont {G.~Z.}\ \bibnamefont
  {Xing}}, \bibinfo {author} {\bibfnamefont {Y.~H.}\ \bibnamefont {Lu}},
  \bibinfo {author} {\bibfnamefont {Y.~F.}\ \bibnamefont {Tian}}, \bibinfo
  {author} {\bibfnamefont {J.~B.}\ \bibnamefont {Yi}}, \bibinfo {author}
  {\bibfnamefont {C.~C.}\ \bibnamefont {Lim}}, \bibinfo {author} {\bibfnamefont
  {Y.~F.}\ \bibnamefont {Li}}, \bibinfo {author} {\bibfnamefont {G.~P.}\
  \bibnamefont {Li}}, \bibinfo {author} {\bibfnamefont {D.~D.}\ \bibnamefont
  {Wang}}, \bibinfo {author} {\bibfnamefont {B.}~\bibnamefont {Yao}}, \bibinfo
  {author} {\bibfnamefont {J.}~\bibnamefont {Ding}}, \ and\ \bibinfo {author}
  {\bibfnamefont {Y.~P.}\ \bibnamefont {Feng}},\ }\href@noop {} {\bibfield
  {journal} {\bibinfo  {journal} {AIP ADVANCES}\ }\textbf {\bibinfo {volume}
  {1}},\ \bibinfo {pages} {022152} (\bibinfo {year} {2011})}\BibitemShut
  {NoStop}%
\bibitem [{\citenamefont {Abraham}\ \emph {et~al.}(2005)\citenamefont
  {Abraham}, \citenamefont {Frank},\ and\ \citenamefont {Guha}}]{abraham05}%
  \BibitemOpen
  \bibfield  {author} {\bibinfo {author} {\bibfnamefont {D.~W.}\ \bibnamefont
  {Abraham}}, \bibinfo {author} {\bibfnamefont {M.~M.}\ \bibnamefont {Frank}},
  \ and\ \bibinfo {author} {\bibfnamefont {S.}~\bibnamefont {Guha}},\
  }\href@noop {} {\bibfield  {journal} {\bibinfo  {journal} {Applied Physics
  Letters}\ }\textbf {\bibinfo {volume} {87}},\ \bibinfo {pages} {252502}
  (\bibinfo {year} {2005})}\BibitemShut {NoStop}%
\bibitem [{\citenamefont {Jagadish}\ and\ \citenamefont
  {Pearton}(2006)}]{jag06}%
  \BibitemOpen
  \bibfield  {author} {\bibinfo {author} {\bibfnamefont {C.}~\bibnamefont
  {Jagadish}}\ and\ \bibinfo {author} {\bibfnamefont {S.~J.}\ \bibnamefont
  {Pearton}},\ }\href@noop {} { {  \textit{Zinc Oxide bulk, thin
  films and nanostructures}}}\ (\bibinfo  {publisher} {Elsevier Press, UK},\
  \bibinfo {year} {2006})\BibitemShut {NoStop}%
\bibitem [{\citenamefont {Sheng}\ \emph {et~al.}(2014)\citenamefont {Sheng},
  \citenamefont {Nakamura}, \citenamefont {Koshibae}, \citenamefont {Makino},
  \citenamefont {Tokura}, ,\ and\ \citenamefont {Kawasaki}}]{Zheng}%
  \BibitemOpen
  \bibfield  {author} {\bibinfo {author} {\bibfnamefont {Z.}~\bibnamefont
  {Sheng}}, \bibinfo {author} {\bibfnamefont {M.}~\bibnamefont {Nakamura}},
  \bibinfo {author} {\bibfnamefont {W.}~\bibnamefont {Koshibae}}, \bibinfo
  {author} {\bibfnamefont {T.}~\bibnamefont {Makino}}, \bibinfo {author}
  {\bibfnamefont {Y.}~\bibnamefont {Tokura}}, , \ and\ \bibinfo {author}
  {\bibfnamefont {M.}~\bibnamefont {Kawasaki}},\ }\href@noop {} {\bibfield
  {journal} {\bibinfo  {journal} {Nature communications}\ }\textbf {\bibinfo
  {volume} {5}},\ \bibinfo {pages} {5584} (\bibinfo {year} {2014})}\BibitemShut
  {NoStop}%
\bibitem [{\citenamefont {Chen}\ \emph {et~al.}(2010)\citenamefont {Chen},
  \citenamefont {Tien},\ and\ \citenamefont {Hsu}}]{Lung}%
  \BibitemOpen
  \bibfield  {author} {\bibinfo {author} {\bibfnamefont {L.-C.}\ \bibnamefont
  {Chen}}, \bibinfo {author} {\bibfnamefont {C.-H.}\ \bibnamefont {Tien}}, \
  and\ \bibinfo {author} {\bibfnamefont {Y.-Y.}\ \bibnamefont {Hsu}},\
  }\href@noop {} {\bibfield  {journal} {\bibinfo  {journal} {Jpn. J. Appl.
  Phys}\ }\textbf {\bibinfo {volume} {49}},\ \bibinfo {pages} {063002}
  (\bibinfo {year} {2010})}\BibitemShut {NoStop}%
\bibitem [{\citenamefont {Zapata}\ \emph {et~al.}(2011)\citenamefont {Zapata},
  \citenamefont {Khalid}, \citenamefont {Simonelli}, \citenamefont
  {Villafuerte}, \citenamefont {Heluani},\ and\ \citenamefont
  {Esquinazi}}]{zap11}%
  \BibitemOpen
  \bibfield  {author} {\bibinfo {author} {\bibfnamefont {C.}~\bibnamefont
  {Zapata}}, \bibinfo {author} {\bibfnamefont {M.}~\bibnamefont {Khalid}},
  \bibinfo {author} {\bibfnamefont {G.}~\bibnamefont {Simonelli}}, \bibinfo
  {author} {\bibfnamefont {M.}~\bibnamefont {Villafuerte}}, \bibinfo {author}
  {\bibfnamefont {S.~P.}\ \bibnamefont {Heluani}}, \ and\ \bibinfo {author}
  {\bibfnamefont {P.}~\bibnamefont {Esquinazi}},\ }\href@noop {} {\bibfield
  {journal} {\bibinfo  {journal} {Appl. Phys. Lett.}\ }\textbf {\bibinfo
  {volume} {99}},\ \bibinfo {pages} {112503} (\bibinfo {year}
  {2011})}\BibitemShut {NoStop}%
\bibitem [{\citenamefont {Cai}\ \emph {et~al.}(2012)\citenamefont {Cai},
  \citenamefont {Wang}, \citenamefont {Yuan},\ and\ \citenamefont
  {Duan}}]{Cai}%
  \BibitemOpen
  \bibfield  {author} {\bibinfo {author} {\bibfnamefont {F.}~\bibnamefont
  {Cai}}, \bibinfo {author} {\bibfnamefont {J.}~\bibnamefont {Wang}}, \bibinfo
  {author} {\bibfnamefont {Z.}~\bibnamefont {Yuan}}, \ and\ \bibinfo {author}
  {\bibfnamefont {Y.}~\bibnamefont {Duan}},\ }\href@noop {} {\bibfield
  {journal} {\bibinfo  {journal} {J. of Power Sources}\ }\textbf {\bibinfo
  {volume} {216}},\ \bibinfo {pages} {269} (\bibinfo {year}
  {2012})}\BibitemShut {NoStop}%
\bibitem [{\citenamefont {Park}\ \emph {et~al.}(2011)\citenamefont {Park},
  \citenamefont {Jo}, \citenamefont {Hong}, \citenamefont {Yoon}, \citenamefont
  {Choe}, \citenamefont {Lee}, \citenamefont {Ji}, \citenamefont {Kim},
  \citenamefont {Kahn}, \citenamefont {Lee}, \citenamefont {Wang},\ and\
  \citenamefont {Lee}}]{Park}%
  \BibitemOpen
  \bibfield  {author} {\bibinfo {author} {\bibfnamefont {W.}~\bibnamefont
  {Park}}, \bibinfo {author} {\bibfnamefont {G.}~\bibnamefont {Jo}}, \bibinfo
  {author} {\bibfnamefont {W.-K.}\ \bibnamefont {Hong}}, \bibinfo {author}
  {\bibfnamefont {J.}~\bibnamefont {Yoon}}, \bibinfo {author} {\bibfnamefont
  {M.}~\bibnamefont {Choe}}, \bibinfo {author} {\bibfnamefont {S.}~\bibnamefont
  {Lee}}, \bibinfo {author} {\bibfnamefont {Y.}~\bibnamefont {Ji}}, \bibinfo
  {author} {\bibfnamefont {G.}~\bibnamefont {Kim}}, \bibinfo {author}
  {\bibfnamefont {Y.~H.}\ \bibnamefont {Kahn}}, \bibinfo {author}
  {\bibfnamefont {K.}~\bibnamefont {Lee}}, \bibinfo {author} {\bibfnamefont
  {D.}~\bibnamefont {Wang}}, \ and\ \bibinfo {author} {\bibfnamefont
  {T.}~\bibnamefont {Lee}},\ }\href@noop {} {\bibfield  {journal} {\bibinfo
  {journal} {Nanotechlogy}\ }\textbf {\bibinfo {volume} {22}},\ \bibinfo
  {pages} {205204} (\bibinfo {year} {2011})}\BibitemShut {NoStop}%
\bibitem [{\citenamefont {Moore}\ and\ \citenamefont
  {Thompson}(2013)}]{MPheno}%
  \BibitemOpen
  \bibfield  {author} {\bibinfo {author} {\bibfnamefont {J.~C.}\ \bibnamefont
  {Moore}}\ and\ \bibinfo {author} {\bibfnamefont {C.~V.}\ \bibnamefont
  {Thompson}},\ }\href@noop {} {\bibfield  {journal} {\bibinfo  {journal}
  {Sensors}\ }\textbf {\bibinfo {volume} {13}},\ \bibinfo {pages} {9921}
  (\bibinfo {year} {2013})}\BibitemShut {NoStop}%
\bibitem [{\citenamefont {Spencer}\ \emph {et~al.}(2013)\citenamefont
  {Spencer}, \citenamefont {Graham}, \citenamefont {Hardman}, \citenamefont
  {Seddon}, \citenamefont {Cliffe}, \citenamefont {Syres}, \citenamefont
  {Thomas}, \citenamefont {Stubbs}, \citenamefont {Sirotti}, \citenamefont
  {Silly}, \citenamefont {Kirkham}, \citenamefont {Kumarasingh}, \citenamefont
  {Hirst}, \citenamefont {Moss}, \citenamefont {Hill}, \citenamefont {Shaw},
  \citenamefont {Chattopadhyay},\ and\ \citenamefont {Flavell}}]{Spencer}%
  \BibitemOpen
  \bibfield  {author} {\bibinfo {author} {\bibfnamefont {B.~F.}\ \bibnamefont
  {Spencer}}, \bibinfo {author} {\bibfnamefont {D.~M.}\ \bibnamefont {Graham}},
  \bibinfo {author} {\bibfnamefont {S.~J.~O.}\ \bibnamefont {Hardman}},
  \bibinfo {author} {\bibfnamefont {E.~A.}\ \bibnamefont {Seddon}}, \bibinfo
  {author} {\bibfnamefont {M.~J.}\ \bibnamefont {Cliffe}}, \bibinfo {author}
  {\bibfnamefont {K.~L.}\ \bibnamefont {Syres}}, \bibinfo {author}
  {\bibfnamefont {A.~G.}\ \bibnamefont {Thomas}}, \bibinfo {author}
  {\bibfnamefont {S.~K.}\ \bibnamefont {Stubbs}}, \bibinfo {author}
  {\bibfnamefont {F.}~\bibnamefont {Sirotti}}, \bibinfo {author} {\bibfnamefont
  {M.~G.}\ \bibnamefont {Silly}}, \bibinfo {author} {\bibfnamefont {P.~F.}\
  \bibnamefont {Kirkham}}, \bibinfo {author} {\bibfnamefont {A.~R.}\
  \bibnamefont {Kumarasingh}}, \bibinfo {author} {\bibfnamefont {G.~J.}\
  \bibnamefont {Hirst}}, \bibinfo {author} {\bibfnamefont {A.~J.}\ \bibnamefont
  {Moss}}, \bibinfo {author} {\bibfnamefont {S.~F.}\ \bibnamefont {Hill}},
  \bibinfo {author} {\bibfnamefont {D.~A.}\ \bibnamefont {Shaw}}, \bibinfo
  {author} {\bibfnamefont {S.}~\bibnamefont {Chattopadhyay}}, \ and\ \bibinfo
  {author} {\bibfnamefont {W.~R.}\ \bibnamefont {Flavell}},\ }\href@noop {}
  {\bibfield  {journal} {\bibinfo  {journal} {Phys. Rev. B}\ }\textbf {\bibinfo
  {volume} {88}},\ \bibinfo {pages} {195301} (\bibinfo {year}
  {2013})}\BibitemShut {NoStop}%
\bibitem [{\citenamefont {Khalid}\ \emph {et~al.}(2009)\citenamefont {Khalid},
  \citenamefont {Ziese}, \citenamefont {Setzer}, \citenamefont {Esquinazi},
  \citenamefont {Lorenz}, \citenamefont {Hochmuth}, \citenamefont {Grundmann},
  \citenamefont {Spemann}, \citenamefont {Butz}, \citenamefont {Brauer},
  \citenamefont {Anwand}, \citenamefont {Fischer}, \citenamefont {Adeagbo},
  \citenamefont {Hergert},\ and\ \citenamefont {Erns}}]{kha09}%
  \BibitemOpen
  \bibfield  {author} {\bibinfo {author} {\bibfnamefont {M.}~\bibnamefont
  {Khalid}}, \bibinfo {author} {\bibfnamefont {M.}~\bibnamefont {Ziese}},
  \bibinfo {author} {\bibfnamefont {A.}~\bibnamefont {Setzer}}, \bibinfo
  {author} {\bibfnamefont {P.}~\bibnamefont {Esquinazi}}, \bibinfo {author}
  {\bibfnamefont {M.}~\bibnamefont {Lorenz}}, \bibinfo {author} {\bibfnamefont
  {H.}~\bibnamefont {Hochmuth}}, \bibinfo {author} {\bibfnamefont
  {M.}~\bibnamefont {Grundmann}}, \bibinfo {author} {\bibfnamefont
  {D.}~\bibnamefont {Spemann}}, \bibinfo {author} {\bibfnamefont
  {T.}~\bibnamefont {Butz}}, \bibinfo {author} {\bibfnamefont {G.}~\bibnamefont
  {Brauer}}, \bibinfo {author} {\bibfnamefont {W.}~\bibnamefont {Anwand}},
  \bibinfo {author} {\bibfnamefont {G.}~\bibnamefont {Fischer}}, \bibinfo
  {author} {\bibfnamefont {W.~A.}\ \bibnamefont {Adeagbo}}, \bibinfo {author}
  {\bibfnamefont {W.}~\bibnamefont {Hergert}}, \ and\ \bibinfo {author}
  {\bibfnamefont {A.}~\bibnamefont {Erns}},\ }\href@noop {} {\bibfield
  {journal} {\bibinfo  {journal} {Phys. Rev. B}\ }\textbf {\bibinfo {volume}
  {80}},\ \bibinfo {pages} {035331} (\bibinfo {year} {2009})}\BibitemShut
  {NoStop}%
\bibitem [{\citenamefont {Khalid}\ \emph {et~al.}(2011)\citenamefont {Khalid},
  \citenamefont {Esquinazi}, \citenamefont {Spemann}, \citenamefont {Anwand},\
  and\ \citenamefont {Brauer}}]{kha11}%
  \BibitemOpen
  \bibfield  {author} {\bibinfo {author} {\bibfnamefont {M.}~\bibnamefont
  {Khalid}}, \bibinfo {author} {\bibfnamefont {P.}~\bibnamefont {Esquinazi}},
  \bibinfo {author} {\bibfnamefont {D.}~\bibnamefont {Spemann}}, \bibinfo
  {author} {\bibfnamefont {W.}~\bibnamefont {Anwand}}, \ and\ \bibinfo {author}
  {\bibfnamefont {G.}~\bibnamefont {Brauer}},\ }\href@noop {} {\bibfield
  {journal} {\bibinfo  {journal} {New Journal of Physics}\ }\textbf {\bibinfo
  {volume} {13}},\ \bibinfo {pages} {063017} (\bibinfo {year}
  {2011})}\BibitemShut {NoStop}%
\bibitem [{\citenamefont {Bridoux}\ \emph {et~al.}(2012)\citenamefont
  {Bridoux}, \citenamefont {Barzola-Quiquia}, \citenamefont {Bern},
  \citenamefont {B\"ohlmann}, \citenamefont {Vrejoiu}, \citenamefont
  {Esquinazi},\ and\ \citenamefont {Ziese}}]{barpat}%
  \BibitemOpen
  \bibfield  {author} {\bibinfo {author} {\bibfnamefont {G.}~\bibnamefont
  {Bridoux}}, \bibinfo {author} {\bibfnamefont {J.}~\bibnamefont
  {Barzola-Quiquia}}, \bibinfo {author} {\bibfnamefont {F.}~\bibnamefont
  {Bern}}, \bibinfo {author} {\bibfnamefont {W.}~\bibnamefont {B\"ohlmann}},
  \bibinfo {author} {\bibfnamefont {I.}~\bibnamefont {Vrejoiu}}, \bibinfo
  {author} {\bibfnamefont {P.}~\bibnamefont {Esquinazi}}, \ and\ \bibinfo
  {author} {\bibfnamefont {M.}~\bibnamefont {Ziese}},\ }\href@noop {}
  {\bibfield  {journal} {\bibinfo  {journal} {Nanotechnology}\ }\textbf
  {\bibinfo {volume} {23}},\ \bibinfo {pages} {085302} (\bibinfo {year}
  {2012})}\BibitemShut {NoStop}%
\bibitem [{\citenamefont {Lorite}\ \emph {et~al.}(2015)\citenamefont {Lorite},
  \citenamefont {Straube}, \citenamefont {Ohldag}, \citenamefont {Kumar},
  \citenamefont {Villafuerte}, \citenamefont {Esquinazi}, \citenamefont
  {Torres}, \citenamefont {de~Heluani}, \citenamefont {Antonov}, \citenamefont
  {Bekenov}, \citenamefont {Ernst}, \citenamefont {Hoffmann}, \citenamefont
  {Nayak}, \citenamefont {Adeagbo}, \citenamefont {Fischer},\ and\
  \citenamefont {Hergert}}]{lor15}%
  \BibitemOpen
  \bibfield  {author} {\bibinfo {author} {\bibfnamefont {I.}~\bibnamefont
  {Lorite}}, \bibinfo {author} {\bibfnamefont {B.}~\bibnamefont {Straube}},
  \bibinfo {author} {\bibfnamefont {H.}~\bibnamefont {Ohldag}}, \bibinfo
  {author} {\bibfnamefont {P.}~\bibnamefont {Kumar}}, \bibinfo {author}
  {\bibfnamefont {M.}~\bibnamefont {Villafuerte}}, \bibinfo {author}
  {\bibfnamefont {P.}~\bibnamefont {Esquinazi}}, \bibinfo {author}
  {\bibfnamefont {R.}~\bibnamefont {Torres}}, \bibinfo {author} {\bibfnamefont
  {S.~P.}\ \bibnamefont {de~Heluani}}, \bibinfo {author} {\bibfnamefont
  {V.~N.}\ \bibnamefont {Antonov}}, \bibinfo {author} {\bibfnamefont {L.~V.}\
  \bibnamefont {Bekenov}}, \bibinfo {author} {\bibfnamefont {A.}~\bibnamefont
  {Ernst}}, \bibinfo {author} {\bibfnamefont {M.}~\bibnamefont {Hoffmann}},
  \bibinfo {author} {\bibfnamefont {S.~K.}\ \bibnamefont {Nayak}}, \bibinfo
  {author} {\bibfnamefont {W.~A.}\ \bibnamefont {Adeagbo}}, \bibinfo {author}
  {\bibfnamefont {G.}~\bibnamefont {Fischer}}, \ and\ \bibinfo {author}
  {\bibfnamefont {W.}~\bibnamefont {Hergert}},\ }\href@noop {} {\bibfield
  {journal} {\bibinfo  {journal} {Appl. Phys. Lett.}\ }\textbf {\bibinfo
  {volume} {106}},\ \bibinfo {pages} {082406} (\bibinfo {year}
  {2015})}\BibitemShut {NoStop}%
\bibitem [{\citenamefont {Hou}\ \emph {et~al.}(2012)\citenamefont {Hou},
  \citenamefont {Dev}, \citenamefont {Frank}, \citenamefont {Rosenauer},\ and\
  \citenamefont {Voss}}]{Hou}%
  \BibitemOpen
  \bibfield  {author} {\bibinfo {author} {\bibfnamefont {D.~C.}\ \bibnamefont
  {Hou}}, \bibinfo {author} {\bibfnamefont {A.}~\bibnamefont {Dev}}, \bibinfo
  {author} {\bibfnamefont {K.}~\bibnamefont {Frank}}, \bibinfo {author}
  {\bibfnamefont {A.}~\bibnamefont {Rosenauer}}, \ and\ \bibinfo {author}
  {\bibfnamefont {T.}~\bibnamefont {Voss}},\ }\href@noop {} {\bibfield
  {journal} {\bibinfo  {journal} {J. Phys. Chem. C}\ }\textbf {\bibinfo
  {volume} {116}},\ \bibinfo {pages} {19604} (\bibinfo {year}
  {2012})}\BibitemShut {NoStop}%
\bibitem [{\citenamefont {Reyes}\ \emph {et~al.}(2011)\citenamefont {Reyes},
  \citenamefont {Ku}, \citenamefont {Duan}, \citenamefont {Xu}, \citenamefont
  {Garfunkel},\ and\ \citenamefont {Lu}}]{Pavel}%
  \BibitemOpen
  \bibfield  {author} {\bibinfo {author} {\bibfnamefont {P.~I.}\ \bibnamefont
  {Reyes}}, \bibinfo {author} {\bibfnamefont {C.-J.}\ \bibnamefont {Ku}},
  \bibinfo {author} {\bibfnamefont {Z.}~\bibnamefont {Duan}}, \bibinfo {author}
  {\bibfnamefont {Y.}~\bibnamefont {Xu}}, \bibinfo {author} {\bibfnamefont
  {E.}~\bibnamefont {Garfunkel}}, \ and\ \bibinfo {author} {\bibfnamefont
  {Y.}~\bibnamefont {Lu}},\ }\href@noop {} {\bibfield  {journal} {\bibinfo
  {journal} {Appl. Phys. Lett.}\ }\textbf {\bibinfo {volume} {101}},\ \bibinfo
  {pages} {031118} (\bibinfo {year} {2011})}\BibitemShut {NoStop}%
\bibitem [{\citenamefont {Klingshirn}\ \emph {et~al.}(2010)\citenamefont
  {Klingshirn}, \citenamefont {Waag}, \citenamefont {Hoffmann},\ and\
  \citenamefont {Geurts}}]{Claus}%
  \BibitemOpen
  \bibfield  {author} {\bibinfo {author} {\bibfnamefont {C.~F.}\ \bibnamefont
  {Klingshirn}}, \bibinfo {author} {\bibfnamefont {A.}~\bibnamefont {Waag}},
  \bibinfo {author} {\bibfnamefont {A.}~\bibnamefont {Hoffmann}}, \ and\
  \bibinfo {author} {\bibfnamefont {J.}~\bibnamefont {Geurts}},\ }\href@noop {}
  { {\bibinfo {title} \textit{Zinc Oxide: From Fundamental Properties Towards
  Novel Applications}}},\ Springer Series in Material Science\ (\bibinfo
  {publisher} {Springer, Heidelberg, Dordrecht, London, New York},\ \bibinfo
  {year} {2010})\BibitemShut {NoStop}%
\bibitem [{\citenamefont {Vece}\ \emph {et~al.}(2011)\citenamefont {Vece},
  \citenamefont {Kolaric}, \citenamefont {Baert}, \citenamefont {Schweitzer},
  \citenamefont {Obradovic}, \citenamefont {Vallée}, \citenamefont {Lievens},\
  and\ \citenamefont {Clays}}]{Vece}%
  \BibitemOpen
  \bibfield  {author} {\bibinfo {author} {\bibfnamefont {M.~D.}\ \bibnamefont
  {Vece}}, \bibinfo {author} {\bibfnamefont {B.}~\bibnamefont {Kolaric}},
  \bibinfo {author} {\bibfnamefont {K.}~\bibnamefont {Baert}}, \bibinfo
  {author} {\bibfnamefont {G.}~\bibnamefont {Schweitzer}}, \bibinfo {author}
  {\bibfnamefont {M.}~\bibnamefont {Obradovic}}, \bibinfo {author}
  {\bibfnamefont {R.~A.~L.}\ \bibnamefont {Vallée}}, \bibinfo {author}
  {\bibfnamefont {P.}~\bibnamefont {Lievens}}, \ and\ \bibinfo {author}
  {\bibfnamefont {K.}~\bibnamefont {Clays}},\ }\href@noop {} {\bibfield
  {journal} {\bibinfo  {journal} {Nanotechlogy}\ }\textbf {\bibinfo {volume}
  {20}},\ \bibinfo {pages} {135203} (\bibinfo {year} {2011})}\BibitemShut
  {NoStop}%
\bibitem [{\citenamefont {Takahashi}\ \emph {et~al.}(1994)\citenamefont
  {Takahashi}, \citenamefont {Kanamori}, \citenamefont {Kondoh}, \citenamefont
  {Minoura},\ and\ \citenamefont {Ohya}}]{Yasu}%
  \BibitemOpen
  \bibfield  {author} {\bibinfo {author} {\bibfnamefont {Y.}~\bibnamefont
  {Takahashi}}, \bibinfo {author} {\bibfnamefont {M.}~\bibnamefont {Kanamori}},
  \bibinfo {author} {\bibfnamefont {A.}~\bibnamefont {Kondoh}}, \bibinfo
  {author} {\bibfnamefont {H.}~\bibnamefont {Minoura}}, \ and\ \bibinfo
  {author} {\bibfnamefont {Y.}~\bibnamefont {Ohya}},\ }\href@noop {} {\bibfield
   {journal} {\bibinfo  {journal} {Jnp. J. Appl. Phys.}\ }\textbf {\bibinfo
  {volume} {33}},\ \bibinfo {pages} {6611} (\bibinfo {year}
  {1994})}\BibitemShut {NoStop}%
\end{thebibliography}

%

\end{document}